\documentclass[onecolumn,amsmath,amssymb]{qm2008_abs}
\usepackage{epsfig}
\usepackage{dcolumn}
\usepackage{bm}
\begin{document}

\title{{\Large Equilibration in Quark Gluon Plasma }}

\bigskip
\bigskip
\author{\large{Santosh K Das}\email{santosh@veccal.ernet.in}, 
Jan-e Alam and Payal Mohanty}
\affiliation{Variable Energy Cyclotron Centre, 1/AF, Bidhan Nagar, Kolkata 700064, India}
\bigskip
\bigskip

\begin{abstract}
\leftskip1.0cm
\rightskip1.0cm
The hydrodynamic expansion rate of quark gluon plasma (QGP) is evaluated 
and compared with the scattering rate of quarks and gluons within the system. 
Partonic scattering rates evaluated within the ambit of  perturbative Quantum 
Choromodynamics (pQCD) are  found to be smaller than the expansion rate 
evaluated with ideal equation of state (EoS) for the QGP. This indicate that during 
the space-time evolution the system remains out of equilibrium. Enhancement of pQCD 
cross sections and a more realistic EoS keep the partons closer to the 
equilibrium.  
\end{abstract}

\maketitle

\section{Introduction}
The main aim of nuclear collisions at the Relativistic Heavy Ion Collider (RHIC) and 
the Large Hadron Collider (LHC) energies is to create a thermalized system of quarks and
gluons - mimicking the situation that prevailed after few microsecond of the 
big bang. The expansion time scale in the microsecond old universe 
is much larger than the interaction time scale of quarks and gluons. 
Consequently  during the evolution the system maintains equilibrium. 
In the case of heavy ion
collisions (HIC), however, these time scales have comparable magnitude. 
For HIC the two pertinent issues for the equilibration
are (i) do the quarks and gluons 
achieve equilibrium and (ii) in case they do,
can the equilibrium be maintained during the evolution ?
or what kinds of EoS and interaction cross sections are 
required for the maintenance of
equilibration. Some of these issues will be addressed in the present work.

\section{Thermal Equilibrium in Heavy Ion Collisions}
To address the first issue the time evolution of
average momentum ($<p_T>$) of the quarks and gluons have been studied  
and compare with its equilibrium value.
In relaxation time approximation it can be shown that the
evolution of the $<p_T>$ is governed by 
the equation:
\begin{equation}
\frac{d<p_T>}{d\tau}=-\frac{<p_T>-<p_{T\mathrm eq}>}{\tau_{\mathrm relax}}
\label{boltzmann}
\end{equation}
$\tau_{\mathrm relax}$ is the relaxation time evaluated 
with pQCD cross section for $\alpha_s=0.3$, $p_{T\mathrm eq}$ 
is the value of $p_T$ in equilibrium. 
The equilibrium time, $\tau_{\mathrm eq}$ is obtained from the condition:
$<p_T>\rightarrow p_{T\mathrm eq}$  as $\tau\rightarrow \tau_{\mathrm eq}$. 
It has been found that the value $\tau_{\mathrm eq}$ depends weakly
on initial value of $<p_T>$.

To address the second issue we assume that the quarks and gluons
produced initially
in thermal equilibrium and check whether it can
maintain the equilibrium during the entire evolution processes by
comparing their scattering rates with the expansion rate of the matter.
The scattering time scales ($\tau_{\mathrm scatt}$) of various partons are evaluated in
pQCD and these time scales are compared with the expansion time scale ($\tau_{\mathrm exp}$). 
For maintenance of thermal equilibrium the following criteria should be satisfied:  
\begin{equation}
\tau_{\mathrm exp}\geq\alpha\tau_{\mathrm scatt}
\label{Eq1}
\end{equation}
where  $\alpha\sim O(1)$, is a constant. 
The criteria given in Eq.~\ref{Eq1} is reverse to the
one used to study the freeze-out of various species 
of particles during the evolution of the early 
universe~\cite{kolbandturner}. Similar condition is used
in heavy ion collisions also to study freeze-out of 
hadrons~\cite{navarra}. 
Physically Eq.~\ref{Eq1} means that the time taken by a partons to
travel a distance of one mean free path by their thermal motion 
they have collectively receded from each other by less than one mean 
free path~\cite{navarra}. 
The $\tau_{\mathrm scatt}$ is determined for  parton types $i$ 
by the expression
\begin{equation}
\tau_{\mathrm scatt}^i=\left(\sum\sigma_{ij}v_{ij}n_j\right)^{-1}
\label{Eq2}
\end{equation}
where $\sigma_{\mathrm ij}$ is the total cross section for  
particles $i$ and $j$, $v_{ij}$ is the relative velocity between
$i$ and $j$ and $n_j$ is the density of the particle type $j$ in 
the medium.

\par
The scattering time scales for partons have been calculated by taking into
account the  following leading order  
processes $gg \rightarrow gg$,
$gg \rightarrow q\overline{q}$, 
$q(\overline{q})g \rightarrow q(\overline{q})g$, 
$qq \rightarrow qq$, 
$q\overline{q} \rightarrow q\overline{q}$ for 
light flavours and gluons~\cite{pqcd}.
Here $q$ stands for light quarks and $g$ denotes
gluons. For evaluating $\tau_{\mathrm scatt}$ for 
heavy quarks ($Q$) the pQCD processes
are taken from ~\cite{combridge}.
The infrared divergence appearing in case of massless particle
exchange in the $t$-channel has been shielded by Debye mass.

The expansion time scale can be defined as:
\begin{equation}
\tau_{\mathrm exp}^{-1}=\frac{1}{\epsilon(\tau,r)}
\frac{d\epsilon(\tau,r)}{d\tau}
\label{Eq3}
\end{equation}
where $\epsilon(\tau,r)$ is the energy density at a proper time 
$\tau$ and radial co-ordinate $r$. $\epsilon(\tau,r)$
is obtained by solving the hydrodynamical equation:
\begin{equation}
\partial_\mu T^{\mu\nu}=0
\label{Eq3}
\end{equation}
with the assumption of boost invariance along longitudinal
direction~\cite{bjorken} and cylindrical symmetry of the system~\cite{hvg}. 
In Eq.~(\ref{Eq3}), 
$T^{\mu\nu}=(\epsilon+P)u^\mu u^\nu-g^{\mu\nu}P$ 
is
the energy momentum tensor, $P$ is the pressure, $u^\mu$ denotes
four velocity and $g^{\mu\nu}$ stands for metric tensor. We consider
a net baryon free QGP here, therefore the baryonic chemical potential ($\mu_B$)
is zero.
\section{Initial conditions and EoS} 
The expansion rates for RHIC and LHC energies have been calculated using  the 
initial conditions, $T_i=400$ MeV, $\tau_i=0.2$ fm  for RHIC which gives
$dN/dy\sim 1100$ and
$T_i=700$ MeV, $\tau_i=0.08$ fm for LHC giving $dN/dy=2100$~\cite{lhc}. 
The initial radial velocity has been
taken as zero.  Two sets 
of equation of state (EoS) have been
used to study the sensitivity of the results on EoS. 
SET-I: In a first order phase transition scenario -
we use the bag model EOS for the QGP phase and for the hadronic
phase all the resonances with mass $\leq 2.5$ GeV have been
considered~\cite{bm}.
and SET-II: The EOS is taken from lattice QCD (lQCD) calculations performed by 
the MILC collaboration~\cite{MILC}.

\section{Results}
\begin{figure}
\begin{center}
\resizebox{0.95\columnwidth}{!}{%
\includegraphics{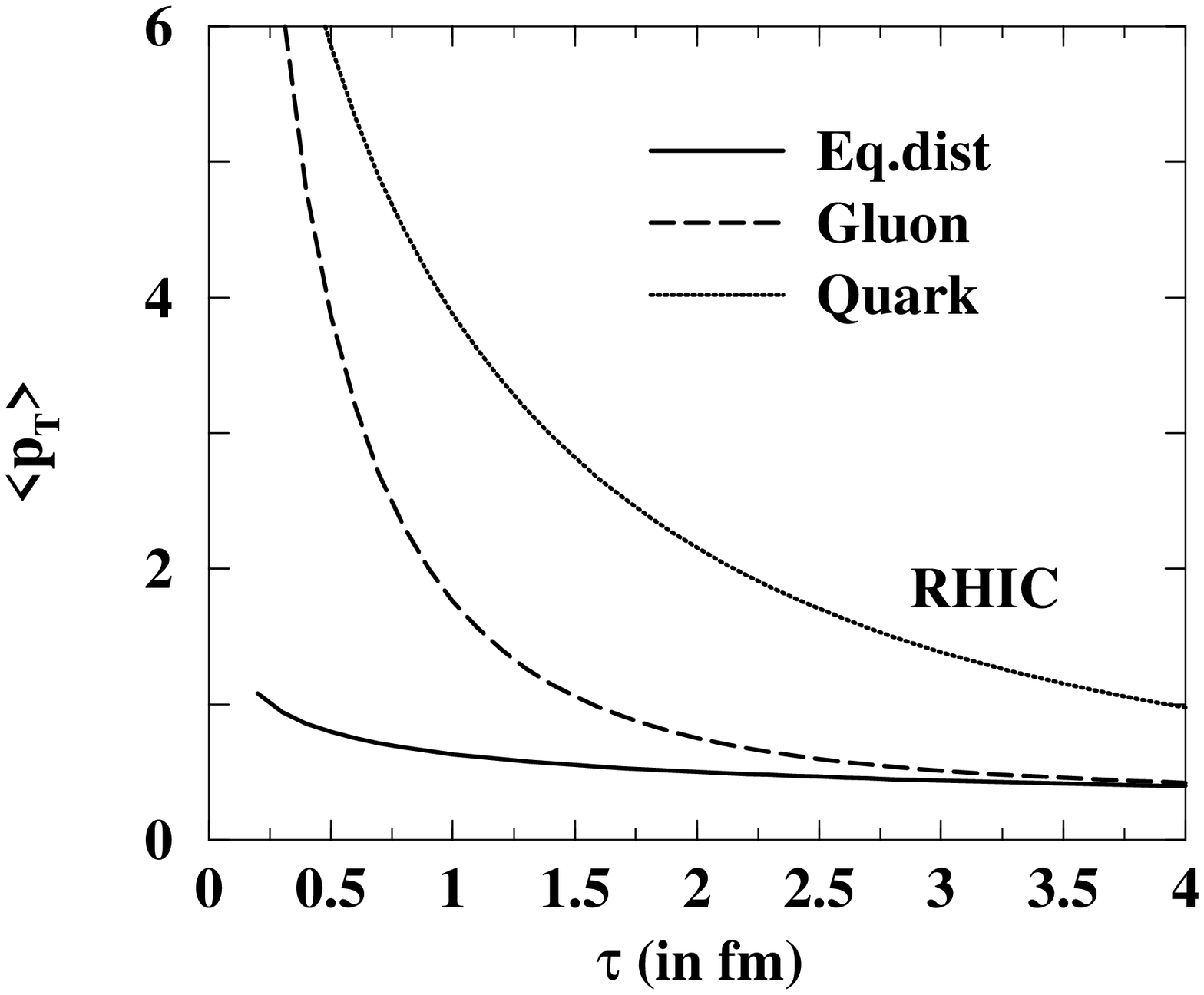}
\includegraphics{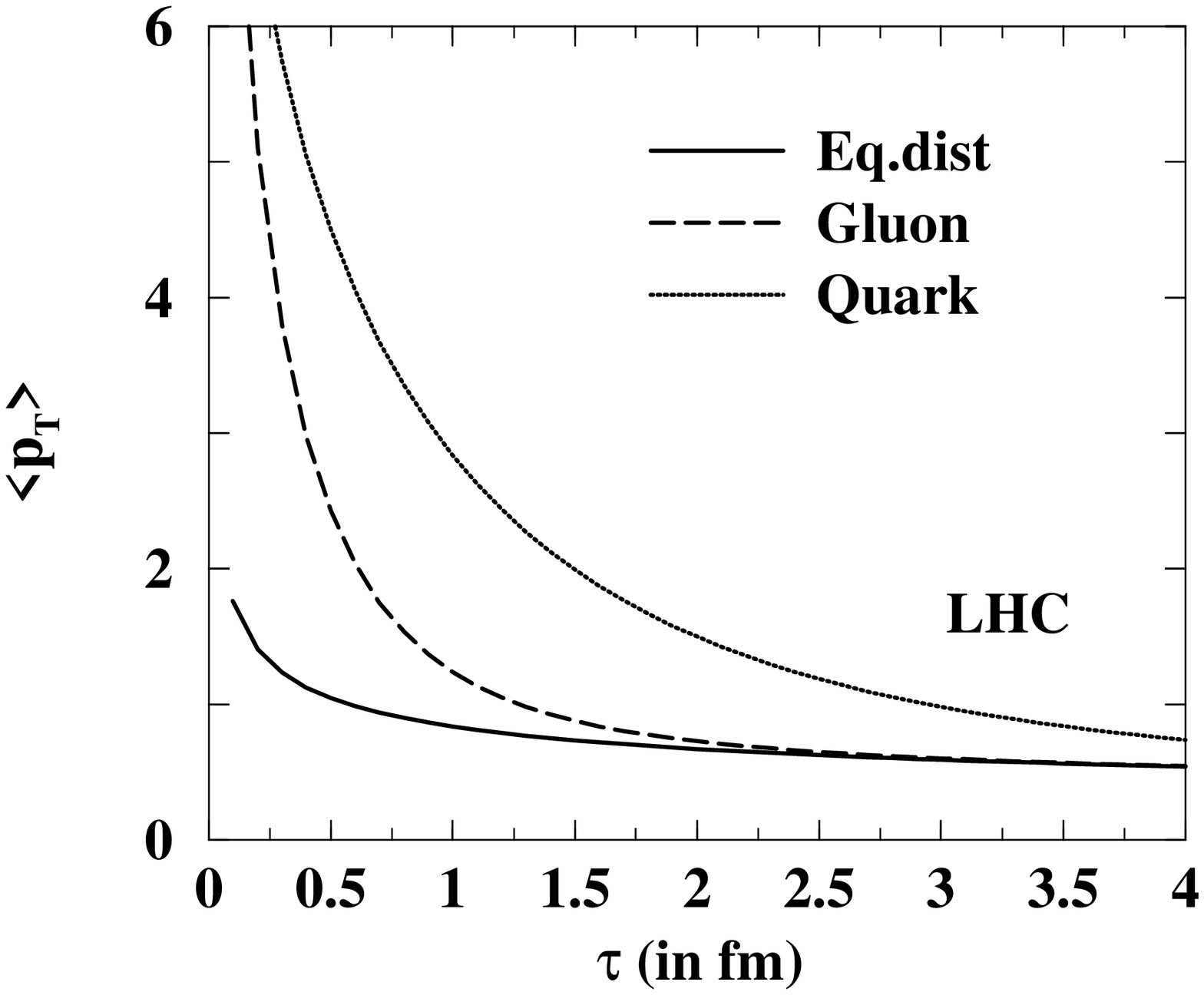}
}
\caption{Evolution of partonic average momentum with time 
for a non-expanding system for RHIC and LHC energies.
}
\label{fig1}
\end{center}
\end{figure}

The time evolution of $<p_T>$ obtained by solving Eq.~\ref{boltzmann} 
is depicted in Fig.~\ref{fig1} for a system of quarks and gluons.
The results with pQCD interactions indicate that the quarks are unlikely 
thermalize both at RHIC and LHC energies. The possibility of gluon 
thermalization at LHC is very strong. The gluon thermalization
time scale seems to comparable to the life time of the QGP at RHIC 
energies indicating that the gluons may remain out of equilibrium 
during the evolution. However, any non-perturbative effects may drive 
the system faster toward equilibrium. 

\begin{figure}
\begin{center}
\resizebox{0.95\columnwidth}{!}{%
\includegraphics{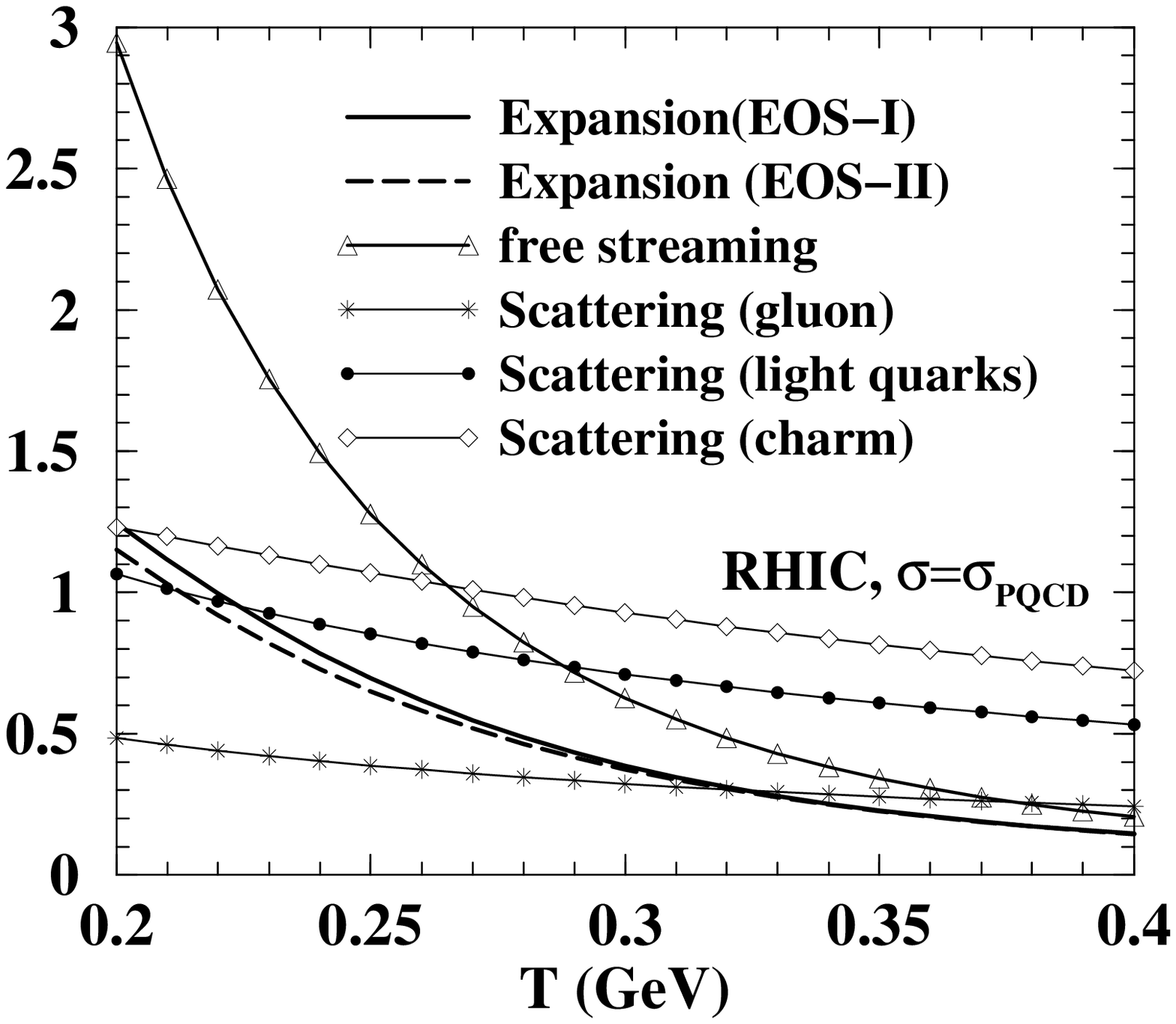}
\includegraphics{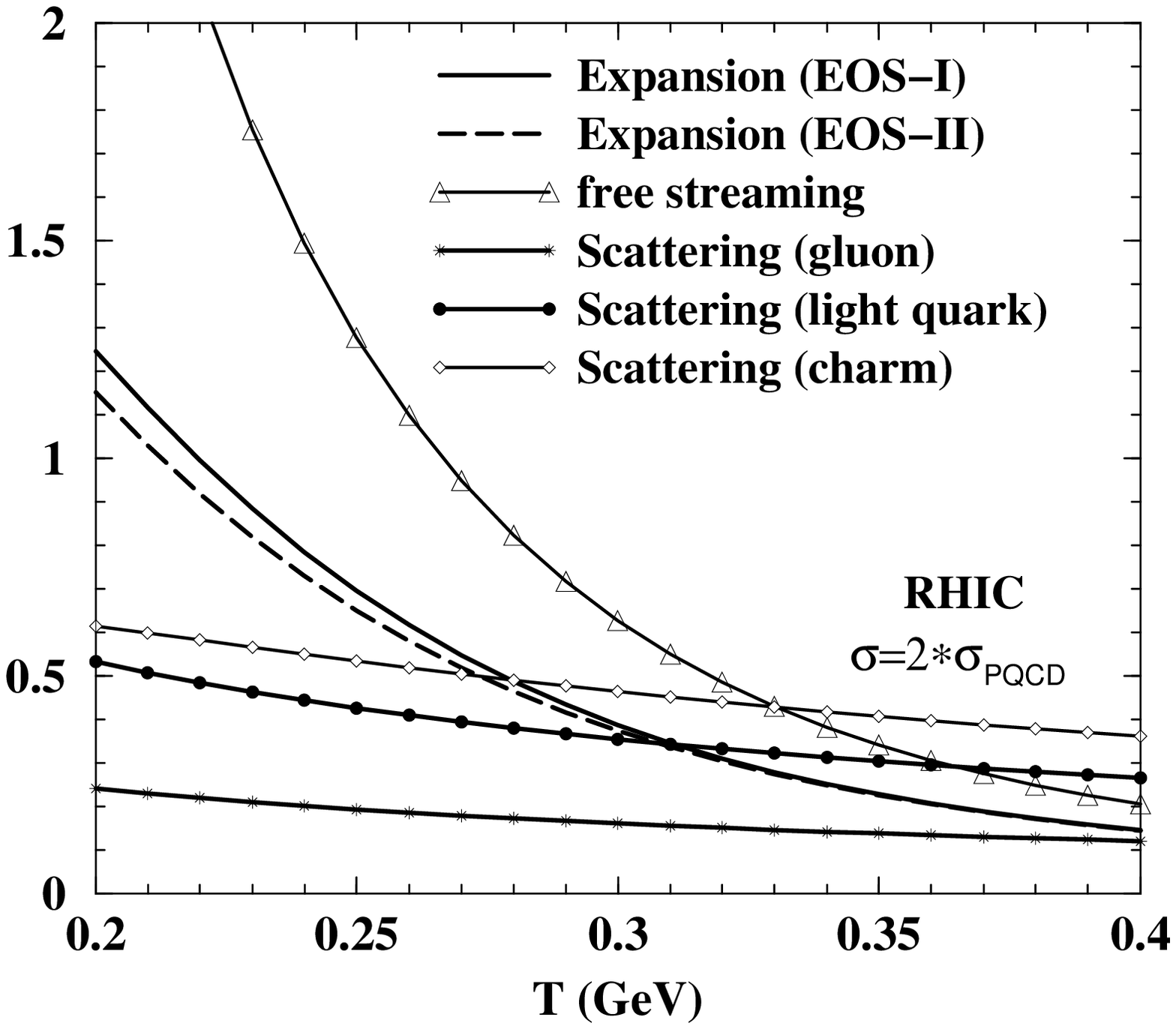}
}
\caption{
Expansion versus scattering time scale for RHIC energy at $r=1$ fm.  
}
\label{fig2}
\end{center}
\end{figure}

In Fig~\ref{fig2} the scattering time scale is contrasted with
the expansion time scale for RHIC energy for two types of EoS 
mentioned above. Expansion rate for both the EoS is similar
at early time but differ little at the late stage of the evolution.
For the sake of comparison the expansion rate for the
extreme case of free streaming
is also displayed. The scattering rates are evaluated with pQCD
cross sections (left panel). The condition for equilibration in Eq.~\ref{Eq1}
indicates that the gluons remains close to equilibrium, however the
charm and bottom (not shown in the figure) quarks remain out of
equilibrium during the entire evolution history.

The analysis of
the experimental data within the ambit of relativistic hydrodynamics
suggest that the matter formed in Au+Au collisions at RHIC achieve
thermalization. One possible reasons for the thermalization to occur
is that the partons interact strongly after their  formation.
It is argued in~\cite{kovchegov} that the onset of thermalization
in the system formed in heavy ion collisions at relativistic
energies can not be achieved without non-perturbative effects.
It has also been shown in~\cite{molnar} that
a large enhancement of the pQCD cross section is required for
the reproduction of experimental data on elliptic flow at RHIC
energies.
Therefore, the pQCD cross sections used to derive
the results shown in Fig.~\ref{fig2} (left panel) should include 
non-perturbative
effects. To implement this we  enhance the pQCD cross sections
by a factor of 2. The resulting scattering time is compared with the
expansion time in Fig.~\ref{fig2} (right panel). It is observed that the gluons
are kept in equilibrium throughout the evolution, light quarks are
closer to the equilibrium as compared to the heavy flavours.

\begin{figure}
\begin{center}
\resizebox{0.95\columnwidth}{!}{%
\includegraphics{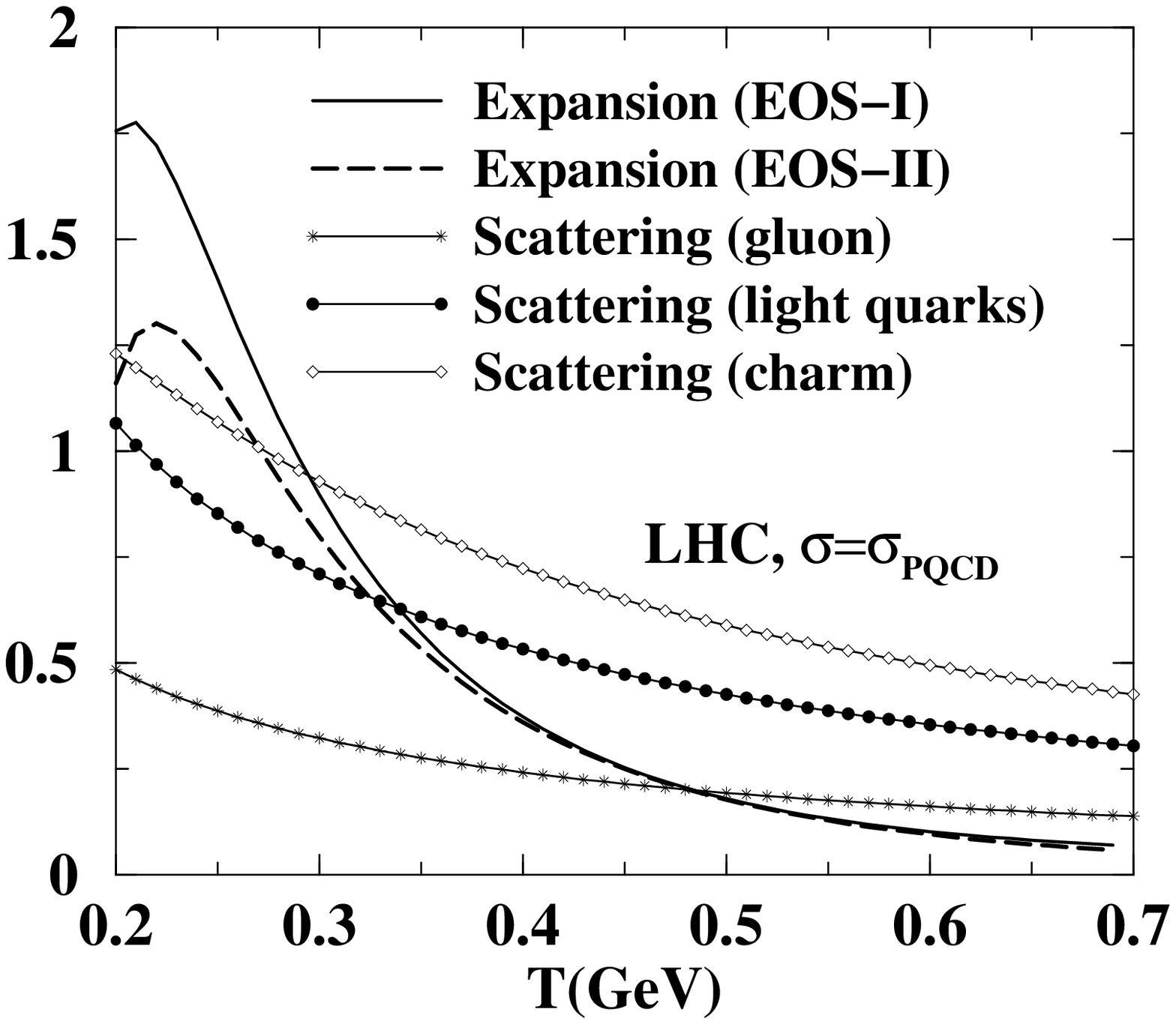}
\includegraphics{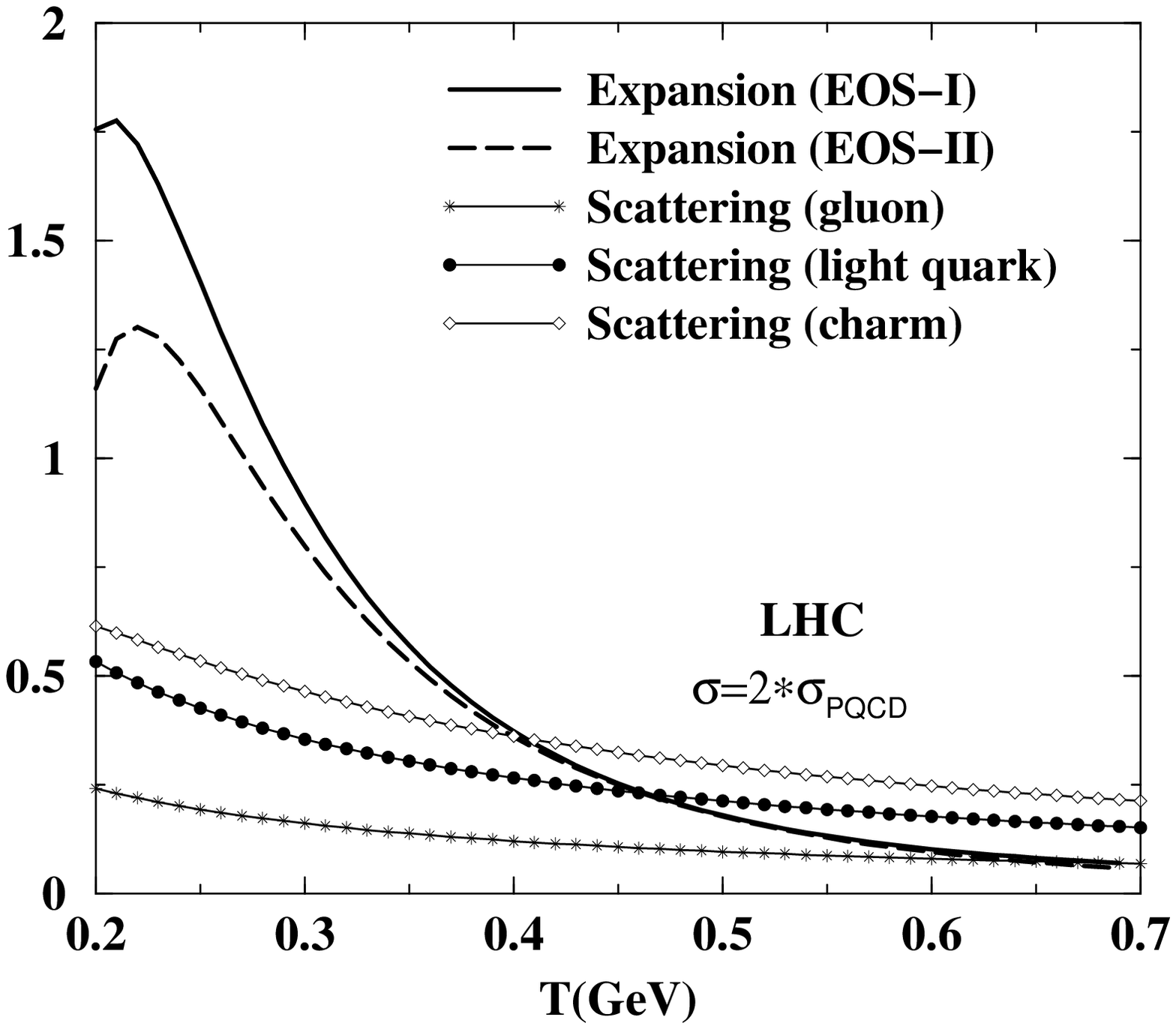}
}
\caption{
Expansion versus scattering time scale for LHC energy at $r=1$ fm.  
}
\label{fig3}
\end{center}
\end{figure}

In Figs.~\ref{fig3}  the results for LHC are displayed
for the two time scales mentioned above for pQCD and enhanced
cross sections. The expansion becomes faster at LHC than RHIC
because of the higher internal pressure. As a consequence,
it is interesting to note that the thermalization scenario at LHC does
not differ drastically from  RHIC.

\section{Summary}
We have evaluated the rate of scattering of various partons  in
QGP and compared the scattering time scale with the expansion
time scale  to examine whether the evolving mater is in thermal equilibrium or 
not.  It is found that gluons remain close to equilibrium throughout the 
evolution. Quarks gets equilibrated at the late stage of the evolution
both for RHIC and LHC energies.  Enhancement of pQCD cross sections and
 realistic lQCD EoS drives the system toward equilibrium faster.

\end{document}